\documentclass[aps,pre,floats,twocolumn,noshowpacs,superscriptaddress]{revtex4-2}

\usepackage{graphicx}
\usepackage{mathtools,amssymb,multirow,rotate}
\usepackage{alphabeta,dcolumn,bm,cancel}

\usepackage{xcolor}
\definecolor{darkviolet}{rgb}{0.58,0.0,0.83}
\definecolor{giocolor}{RGB}{0,150,100}
\definecolor{darkorange}{rgb}{1.0,0.39,0.0}

\usepackage{hyperref}
\hypersetup{colorlinks=true,allcolors=blue}
\usepackage{orcidlink}

\usepackage{times}
\usepackage[T1]{fontenc}

\usepackage[most]{tcolorbox}
\usepackage[svgnames]{xcolor}

\usepackage{ragged2e}

\usepackage{caption}

\definecolor{Header_color}{HTML}{72bea6} 
\definecolor{Body_color}{HTML}{72bea6}

\begin{document}

\title{\textbf{Higher-order interactions in ecology can be hidden in plain sight}}

\author{Violeta Calleja-Solanas\,\orcidlink{0000-0001-7917-8984}}
\affiliation{Department of Biology, University of Oxford, Life and Mind Building, South Parks Road, Oxford OX1 3EL, United Kingdom}
\affiliation{Department of Ecology and Evolution, Doñana Biological Station (EBD-CSIC), 41092 Seville, Spain}

\author{Santiago Lamata-Ot\'in\,\orcidlink{0009-0004-0247-4792}}
\affiliation{Department of Condensed Matter Physics, University of Zaragoza, 50009 Zaragoza, Spain}
\affiliation{GOTHAM lab, Institute of Biocomputation and Physics of Complex Systems (BIFI), University of Zaragoza, 50018 Zaragoza, Spain}

\author{\linebreak Carlos G\'omez-Ambrosi\,\orcidlink{0000-0001-7951-380X}}
\affiliation{Department of Mathematics, University of Zaragoza, 50009 Zaragoza, Spain}
\affiliation{GOTHAM lab, Institute of Biocomputation and Physics of Complex Systems (BIFI), University of Zaragoza, 50018 Zaragoza, Spain}

\author{Jes\'us G\'omez-Garde\~nes\,\orcidlink{0000-0001-5204-1937}}
\affiliation{Department of Condensed Matter Physics, University of Zaragoza, 50009 Zaragoza, Spain}
\affiliation{GOTHAM lab, Institute of Biocomputation and Physics of Complex Systems (BIFI), University of Zaragoza, 50018 Zaragoza, Spain}


\author{Sandro Meloni\,\orcidlink{0000-0001-6202-3302}}
\email{sandro@ifisc.uib-csic.es}
\affiliation{Institute for Cross-Disciplinary Physics and Complex Systems (IFISC), CSIC-UIB, 07122 Palma de Mallorca, Spain}
\affiliation{Centro Studi e Ricerche ``Enrico Fermi" (CREF), 00184 Rome, Italy}

\date{\today}

\begin{abstract}
Higher-order interactions are increasingly recognized as a key component of ecological dynamics. However, we show that higher-order Lotka–Volterra dynamics can, in some scenarios, be accurately reproduced by effective pairwise models fitted to the same abundance time series. Consequently, higher-order interactions cannot, in general, be inferred from time-series data alone. We further identify a fundamental problem of mechanistic identifiability, whereby distinct interaction mechanisms generate nearly indistinguishable dynamics, potentially leading to accurate yet misleading ecological interpretations. Our results highlight the need to complement time-series data with additional ecological information to infer interaction structure reliably.
\end{abstract}

\maketitle

\noindent \textbf{\large Introduction}

\noindent Higher-order interactions (HOIs) in ecology attracted much attention in ecological research several decades ago, in part because their nonlinearity promised greater realism than the classical pairwise Lotka–Volterra equations \cite{abrams_arguments_1983,billick_higher_1994}. After their popularity waned amid debates over their definition \cite{pomerantz_higher_1981,wootton_putting_1994} and their inference \cite{case_testing_1981,yodzis_indeterminacy_1988}, HOIs are currently enjoying a revival \cite{,grilli_higher-order_2017, battiston_physics_2021,gibbs2024can}. The renewed interest is reflected in empirical and semi-empirical studies that increasingly incorporate HOI terms into phenomenological models fit to data \cite{buche_continuum_2025,buche2024multitrophic}. The original quarrels about their definition and interpretation in nature are slowly settling. Higher-order interactions broadly refer to non-additive effects of density on per capita growth \cite{letten_mechanistic_2019}. HOIs derive from mechanisms that exist only when three or more species are present. Then, group interactions, multivariate interactions, and interaction modifications fall under the umbrella of HOIS. \cite{kleinhesselink_detecting_2022, sanchez2019defining}. In parallel, recent theoretical work has clarified that HOIs can stabilize dynamics in multispecies communities \cite{bairey_high-order_2016,gibbs_coexistence_2022, grilli_higher-order_2017} and be a predictor of community dynamics \cite{mayfield_higher-order_2017}.

However, the rationale for adding HOIs to models used to fit data is rarely justified, and doing so moves us away from parsimonious descriptions based on a minimal set of ecologically meaningful parameters \cite{aho2014model,aho2016foundational}. Even though Nature rarely operates in an additive fashion \cite{strogatz2015}, our models should be based on a set of parameters identified by the researchers as meaningful, including, when necessary, covariates and higher-order terms, but with as few parameters as possible (being as simple as possible, but no simpler). Critically, when comparing a more complex model with HOIs to an effective pairwise one, the inclusion of higher-order terms almost always improves explanatory power in a statistical sense \cite{aladwani_ecological_2020}. Any claimed biological role for a higher-order mechanism must therefore be weighed against the baseline expectation that additional flexibility improves predictive power even in the absence of true HO mechanisms \cite{kleinhesselink_detecting_2022,lai2024detecting}.  In practice, HOIs are often inferred by extending generalized Lotka–Volterra equations with higher-order terms and fitting them to time series of species abundances \cite{ludington2022higher, davis2022methods, malizia_reconstructing_2023, mickalide2019higher}, or fitting to a model of interaction data/neighborhoods \cite{buche_continuum_2025}. The key difficulty is that standard tools of model comparison and goodness of fit are silent about mechanistic identifiability: they indicate which model fits better, but not whether HOIs are genuinely required or simply absorbed into effective pairwise interactions that yield almost-indistinguishable dynamics.

Implicitly, these fittings assume that given sufficiently rich data, we can decide whether the underlying dynamics is fundamentally pairwise or genuinely higher-order. However, this assumption has rarely been examined in its most favorable light. Suppose we are in an idealized setting with high-resolution abundance trajectories, no observation error, and full control over the functional form of the model. Under such conditions, one might expect identifiability to be limited only by data quantity or by the magnitude of HOI effects. 

Here we ask the disquieting question: even under these idealized conditions, can the presence or absence of HOIs always be deduced from abundance dynamics alone? To address this question, we developed a systematic computational pipeline to test whether higher-order interactions leave a unique, detectable dynamical signature. Using second-order Lotka-Volterra systems as our general framework, we generated in silico, high-resolution time series of species abundances for communities with known higher-order interactions. We then inferred a standard pairwise Lotka-Volterra model from the same data by evaluating the per capita growth rates from these trajectories and fitting them to a hyperplane. Finally, we integrated this pairwise system and compared its resulting dynamics, attractors, and interaction coefficients against those of the original higher-order model to determine if the underlying mechanisms could be distinguished. 

We show that the answer to our question is no: even under ideal observation conditions, the presence of HOIs cannot always be deduced from abundance dynamics alone. We constructed scenarios in which dynamics with substantial HOIs can be approximated with striking accuracy by a pairwise model to the same time series. Furthermore, the inferred pairwise model not only reproduces the observed trajectories but also assigns qualitatively different ecological roles to species, reversing the direction or sign of their interactions. These phenomena reflect a fundamental problem of structural non-identifiability: along a given trajectory, higher-order contributions can be flattened into effective pairwise coefficients. Consequently, HOIs are practically invisible to time-series inference, meaning standard pairwise fits can yield compelling but mechanistically different ecological interpretations, making the presence of HOIs intrinsically model-dependent rather than empirically decidable \cite{lucas2024reducibility}. The practical cost of this invisibility is highest precisely when we care most about the consequences of interventions, biological invasions, climate anomalies, or management actions —because models may fail when the system is forced into regions of state space where higher-order effects become decisive.

\begin{figure}[t] 
    \centering
    \includegraphics[width = 0.88\linewidth]{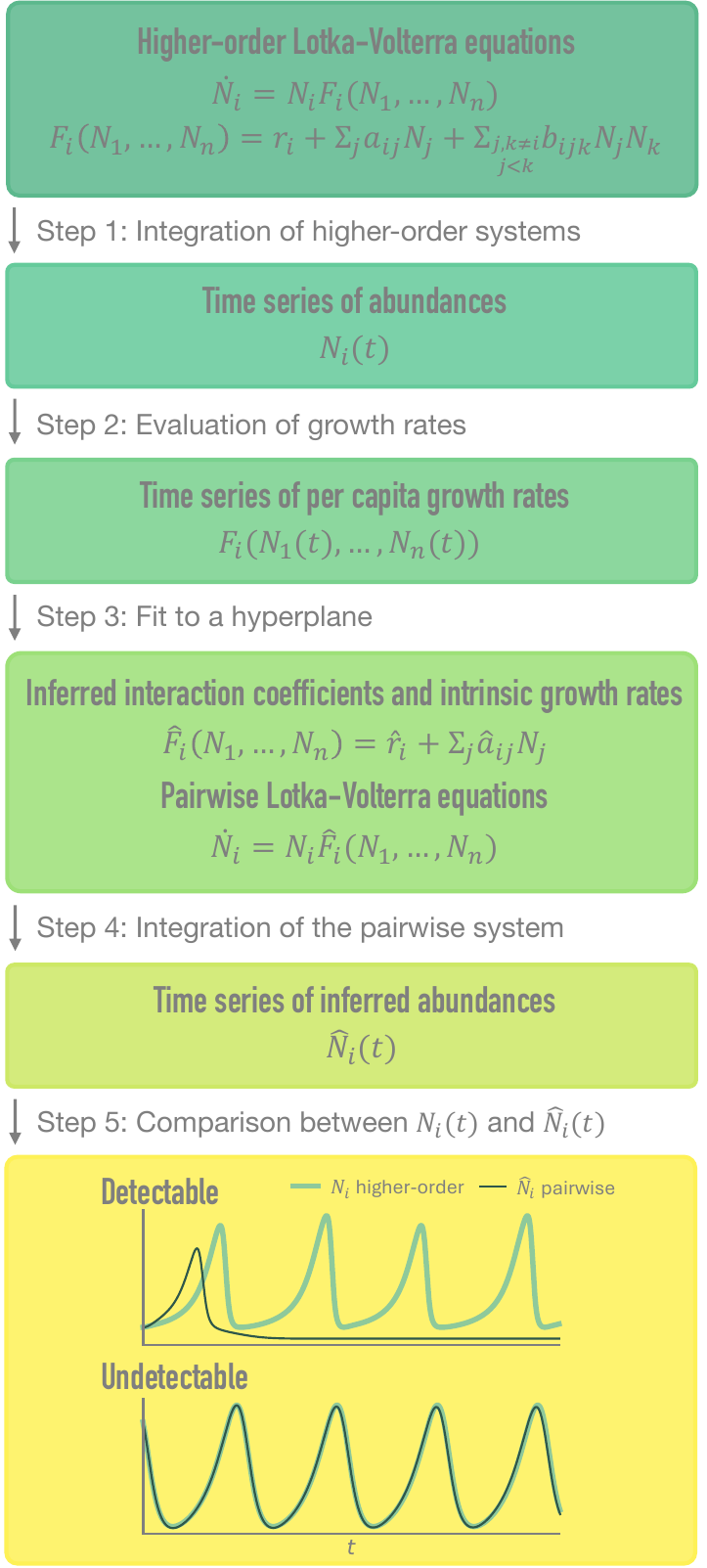}
    \caption{\justifying \textbf{Workflow to assess whether higher-order interactions are detectable from abundance time series.} From the time series of abundances obtained by integrating the original higher-order system, we evaluate the per capita growth rates, fit them to a hyperplane to infer interaction coefficients and intrinsic growth rates, and integrate the resulting pairwise system. The final comparison between the inferred and original dynamics reveals whether higher-order interactions are detectable.}
    \label{fig:1}
\end{figure}
\bigskip

\noindent \textbf{\large Material and methods} \medskip

\noindent \textbf{Pairwise and higher-order Lotka-Volterra models} \newline
We begin by recalling the pairwise (generalized) and higher-order Lotka-Volterra models, and then explain how to obtain a pairwise Lotka-Volterra model from a time series of species abundances generated by a higher-order Lotka-Volterra model.

We consider a population of $n$ interacting species whose dynamics is governed by a system of ecological (or Kolmogorov) equations~\cite{sigmund2007,patel2018} of the form
\begin{equation} \label{eq:Kolmogorov}
    \dot N_i = N_i F_i (N_1, \ldots, N_n), \quad i = 1, \ldots, n,
\end{equation}
where $N_i$ is the abundance of species $i$, $\dot N_i$ its time derivative, i.e.\ the growth rate of species $i$, and $F_i$ a sufficiently smooth function that encodes the per capita growth rate ($\dot N_i / N_i$) of species $i$.
In the simplest case of the generalized Lotka-Volterra (gLV) equations,
\begin{equation} \label{eq:gLV}
    F_i (N_1,\ldots,N_n) = r_i + \sum_j a_{ij} N_j,
\end{equation}
where $r_i$ is the intrinsic growth rate of species $i$, and $a_{ij}$ measures the direct effect of species $j$ on the per capita growth rate of species $i$.
The intra-specific coefficient $a_{ii}$ accounts for the self-regulation of species $i$, whereas the inter-specific coefficients $a_{ij}$, for $j \not= i$, account for the pairwise interactions between species.
Depending on the signs of $a_{ij}$ and $a_{ji}$, these interactions are interpreted as predator-prey $(+\, -)$, competitive $(- \, -)$, mutualistic $(+ \, +)$, etc.
The matrix $A = (a_{ij})$ is the interaction matrix of the system, and the vector $\mathbf{r} = (r_1, \ldots, r_n)$ is the vector of intrinsic growth rates.
Note that in Eq. (\ref{eq:gLV}) pairwise interactions are codified by linear functions $F_i$.

\begin{figure*}
    \centering
    \includegraphics[width = 1.\linewidth]{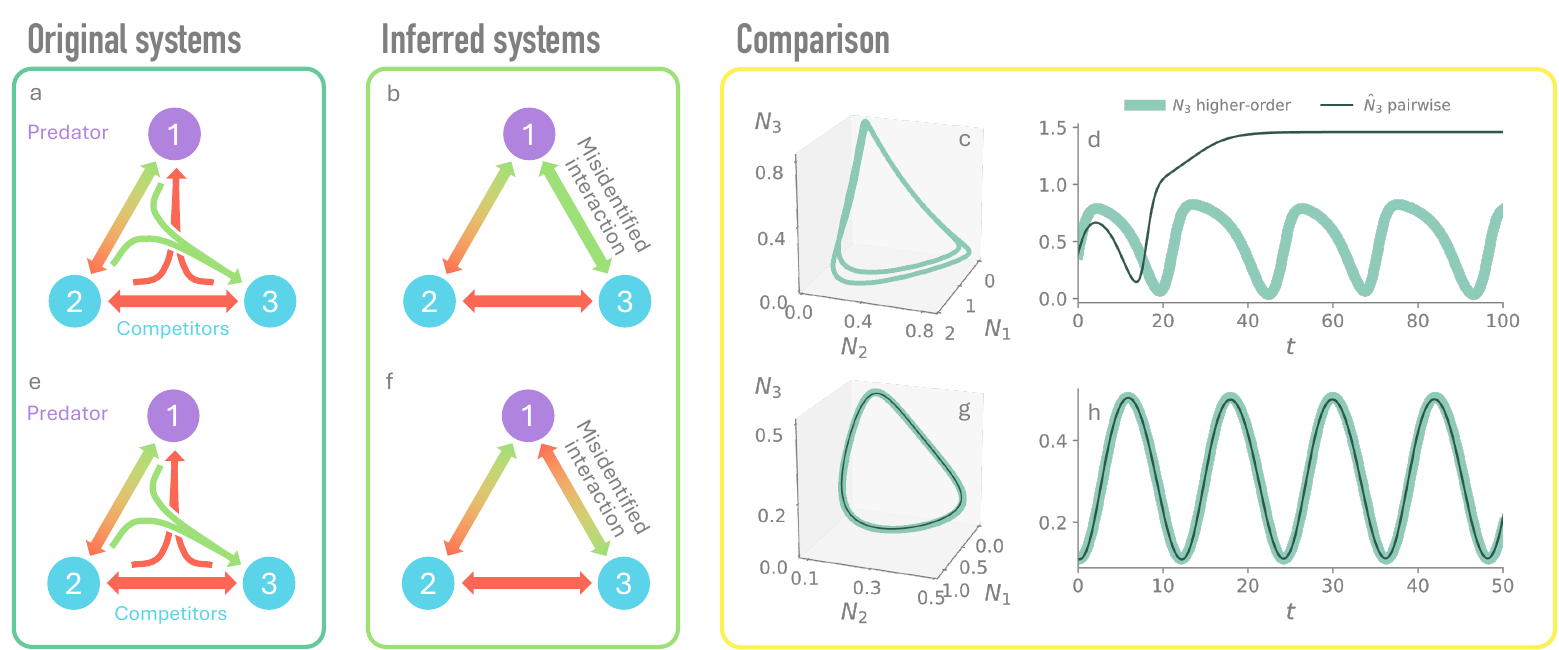}
\caption{\justifying
\textbf{Examples of detectable and undetectable higher-order interactions.}
Panels \textbf{a}–\textbf{d} correspond to System 1 (detectable), and panels \textbf{e}–\textbf{h} to System 2 (undetectable).
Panels \textbf{a} and \textbf{e} show the original interaction structure, including higher-order interactions (join arrows), while
panels \textbf{b} and \textbf{f} display the inferred pairwise interaction networks obtained from the fitting procedure. In these network representations, green (red) arrows indicate positive (negative) interaction coefficients.
Panels \textbf{c} and \textbf{g} compare the original and the inferred attractors, and panels \textbf{d} and \textbf{h} compare the corresponding time series of species 3. In System 1, higher-order interactions leave dynamical signatures that cannot be captured by the pairwise approximation. In contrast, for System 2, the pairwise approximation reproduces the dynamics ($\rho_i \approx 0.99$ for all species).}
    \label{fig:Fig2}
\end{figure*}

The next step in complexity consists of adding quadratic terms of the form $b_{ijk} N_j N_k$ to the functions $F_i$, so that they become polynomial functions of degree 2.
If $i$, $j$, and $k$ are distinct, $b_{ijk}$ measures the joint direct effect of species $j$ and $k$ on the per capita growth rate of species $i$.
Thus, the coefficients $b_{ijk}$ account for the strength of the higher-order (triadic in this case) interactions among species.
Since we can group the terms $b_{ijk} N_j N_k$ and $b_{ikj} N_k N_j$ together, we can assume, without loss of generality, that $b_{ikj} = 0$ if $k > j$.
Moreover, for simplicity and interpretability, we will assume that the coefficients $b_{ijk} = 0$ whenever $i$, $j$, and $k$ are not distinct.
In the sense of \cite{kleinhesselink_detecting_2022}, we retain the higher-order interactions that are \textit{hard}, and we discard those that are \textit{soft} (such as $b_{iij} N_i N_j$) or that are not considered higher-order (such as $b_{iii} N_i^2$ or $b_{ijj} N_j^2$).
We arrive in this way at a second-order special case of the higher-order Lotka-Volterra (hoLV) equations:
\begin{equation} \label{eq:hoLV}
    F_i (N_1,\ldots,N_n) = r_i + \sum_j a_{ij} N_j + \sum_{
    \scriptstyle j,k \not= i \atop \scriptstyle j<k
    } b_{ijk} N_j N_k.
\end{equation}

We could complicate matters further by adding terms of order 3 or higher to the functions $F_i$. Eqs.~\eqref{eq:gLV} and \eqref{eq:hoLV} can be interpreted as containing terms up to order 1 or up to order 2, respectively, in the Taylor series expansion of more complex functions $F_i$ \cite{lotka1956,macarthur1970,may1973,letten2019}. In contrast to the previous paragraph, the important thing to note is that higher-order interactions are now codified by non-linear functions $F_i$.
\medskip

\noindent \textbf{Relative weight of higher-order interactions} \newline
To quantify the relevance of higher-order terms in Eqs.~\eqref{eq:hoLV}, we define the relative weight of higher-order interactions as
\begin{equation} \label{eq:relative_weight}
    \theta = \frac{\sum_{i,j,k}|b_{ijk}|}{\sum_{i,j}|a_{ij}|+\sum_{i,j,k}|b_{ijk}|}.
\end{equation}

\begin{figure*}[t]
    \centering
    \includegraphics[width = 1.\linewidth]{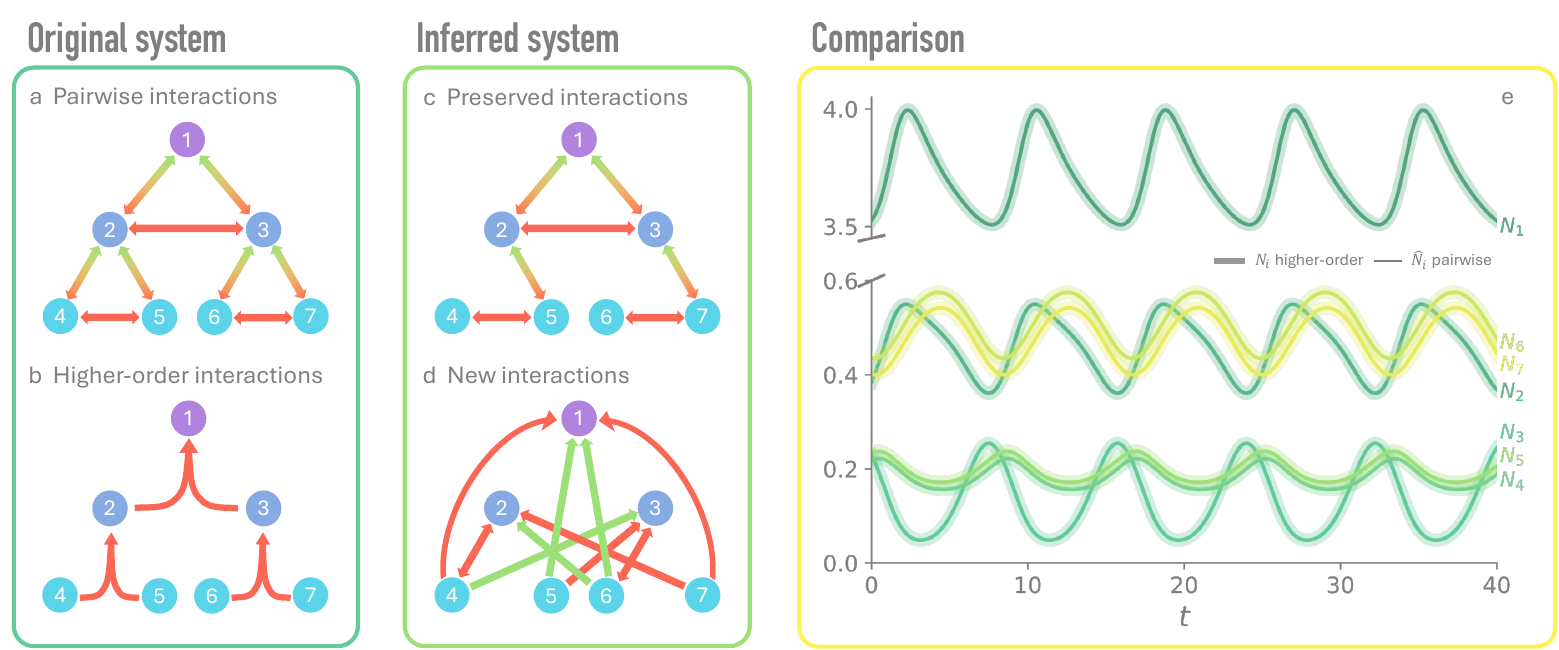}
    \caption{\justifying \textbf{Undetectable higher-order interactions in a multispecies system.}
    Panels \textbf{a} and \textbf{b} show the original pairwise and higher-order interaction structure.
    The inferred pairwise interaction network (panels \textbf{c} and \textbf{d}) present two misidentified interactions (between species 2 and 4, and species 3 and 6), as well as new spurious interactions. Green (red) arrows indicate positive (negative) interaction coefficients. Panel \textbf{e} shows the comparison between the original and inferred time series of species abundances.  The pairwise approximation closely reproduces the higher-order dynamics, with nearly identical trajectories for all species, but introduces new interactions that alter the ecological interpretation of the system.}
    \label{fig:Fig3}
\end{figure*}
\medskip

\noindent \textbf{Pairwise approximation} \newline
In Steps 1--5, we explain how to, based on the non-linear structure of HOIs, fit a pairwise Lotka-Volterra model to a time series of species abundances generated by a higher-order Lotka-Volterra model, and how to interpret the results (see Fig. \ref{fig:1} and Supporting Information for more details) 
\medskip

\noindent \textit{Step 1.—}
Our starting point is a system of higher-order Lotka-Volterra equations (Eqs.~\eqref{eq:Kolmogorov} and \eqref{eq:hoLV}) that we numerically integrate to obtain a time series of abundances $N_i(t)$ for each species $i$, where $t \in \{ 0, 1, \ldots, T \}$.
We assume that there is a time value $T_0 < T$ such that all time series $N_i(t)$ can be considered to have transient behavior for $t \in \{ 0, \ldots, T_0 - 1 \}$, and to have settled on the attractor of the dynamics for $t \in \{ T_0, \ldots, T \}$.
We also assume that there is a time value $s < T - T_0$ such that all time series, for $t \in \{ T_0, \ldots, T_0 + s \}$, reflect the short-term behavior of the system.
The intended interpretation is that the ecological dynamics lie on the attractor and that transient behavior is merely an artifact of the model equations.
\medskip

\noindent\textit{Step 2.—}
We compute the time series of per capita growth rates $F_i(N_1(t), \ldots, N_n(t))$, for $t \in \{ T_0, \ldots, T \}$ (i.e.\ for points in the attractor of the higher-order dynamics), and obtain a cloud of points $(N_1(t), \ldots, N_n(t), F_i(N_1(t), \ldots, N_n(t)))$ in $(n+1)$-dimensional space, all of them belonging to the graph of the function $F_i (N_1, \ldots, N_n)$.
\medskip

\noindent \textit{Step 3.—}
We then fit these points to an $n$-dimensional hyperplane using linear least-squares regression, to infer the linear function $\hat F_i$ that best approximates $F_i$ on the points in the attractor:
\begin{equation} \label{eq:hats}
    \hat F_i (N_1,\ldots,N_n) = \hat r_i + \sum_j \hat a_{ij} N_j.
\end{equation}
This function contains the inferred interaction coefficients and intrinsic growth rates of the pairwise approximation.
\medskip

\noindent \textit{Step 4.—}
We numerically integrate the pairwise gLV equations
\begin{equation} \label{eq:approx}
    \dot N_i = N_i (\hat r_i + \sum_{j} \hat a_{ij} N_j),
\end{equation}
using the inferred coefficients and starting from the initial conditions $\hat N_i(0) = N_i(T_0)$.  We obtain a time series of inferred abundances $\hat N_i(t)$ for each species $i$, where $t \in \{ 0, \ldots, T \}$.
\medskip

\noindent \textit{Step 5.—}
In order to contrast the short-term behavior of the higher-order and pairwise dynamics, we compare the time series $N_i(t)$, for $t \in \{ T_0, \ldots, T_0 + s \}$, with the time series $\hat N_i(t)$, for $t \in \{ 0, \ldots, s \}$.
Beyond visual comparison, we compute the sample Pearson correlation coefficient $\rho_i$ between $N_i$ and $\hat N_i$ as a measure of the accuracy of the short-term pairwise approximation.
On the other hand, we contrast the long-term behavior of the higher-order and pairwise dynamics by comparing their respective attractors: $N_i(t)$ and $\hat N_i(t)$, for $t \in \{ T_0, \ldots, T \}$.
Finally, we compare the ecological parameters (interaction coefficients and intrinsic growth rates) of the inferred pairwise approximation with those of the original higher-order model.

If, instead of being in an ideal in silico scenario, we were given an experimental time series of species abundances $N_i(t)$, for $t \in \{ T_0, \ldots, T \}$, in Step 1, then the main challenge would be how to approximate, for each species $i$, the corresponding time series of time derivatives (i.e.\ growth rates) $\dot N_i(t)$.
Several tools could be used, from neural networks (as developed in \cite{latremoliere2022}, where the Jacobian matrix of an unknown multivariate function is estimated from sample values) to gradient functions.
Dividing then by the species abundances $N_i(t)$, one would get the time series of per capita growth rates, $\dot N_i(t) / N_i(t)$, which would provide an approximation to the missing $F_i(N_1(t), \ldots, N_n(t))$, ready for Step 2.
\bigskip

\noindent \textbf{\large Results}

\noindent We show how the analysis described in the preceding section can yield different outcomes depending on the system under study. To begin with, we consider a community with one predator (species 1), one prey (species 2), and a third species (species 3) that competes with the prey \cite{vandermeer1998}.
The interaction matrix is
\begin{equation} \label{eq:Vandermeer}
    A =
    \left( \begin{array}{rrr}
        0 & 4 & 0 \\
        -1 & -1 & -1 \\
        0 & a_{32} & -1
    \end{array} \right),
\end{equation}
where we initially set $a_{32} = -2$, and the vector of intrinsic growth rates is $\mathbf{r} = (-1,1,1)$. Under the initial conditions $N = (0.1,0.1,0.1)$, the pairwise dynamics stabilizes to a limit cycle. We then introduce higher-order coefficients $(b_{123},b_{213},b_{312}) = (-\beta,0,\beta)$. For $\beta > 0$, the predation of species 1 on species 2 benefits species 3 (positive $b_{312}$), whereas the competition between species 2 and 3 harms species 1 (negative $b_{123}$).

In Fig. \ref{fig:Fig2}a-d, we compare the original system with $\beta=1.1$ (System 1; see the scheme in Fig. \ref{fig:Fig2}a) with the corresponding pairwise approximation (see the scheme in Fig. \ref{fig:Fig2}b). Although the relative weight of higher-order interactions is modest, $\theta \approx 0.18$, their effects are clearly detectable. This can be seen by comparing the attractors shown in Fig.~\ref{fig:Fig2}c. The original higher-order dynamics stabilizes to a limit cycle, whereas the pairwise approximation converges to an equilibrium point. In this equilibrium, $N^*_3\approx1.46$, while species 1 and 2 become extinct (see Fig.~\ref{fig:Fig2}d for the comparison between the time series of species 3).

We now vary the competitive impact of species 2 on species 3 by setting $a_{32} = -3$. In the absence of higher-order terms, the dynamics converges to the equilibrium point $N^* = (0.5,0.25,0.25)$. However, including higher-order terms as before with $\beta = 1.2$ (System 2) changes the nature of the stable attractor from a fixed point to a limit cycle. Note that the relative weight of higher-order interactions remains $\theta\approx0.18$. Remarkably, as shown in Fig. \ref{fig:Fig2}e-h, the higher-order interactions become undetectable: the original higher-order dynamics can be approximated with great accuracy by the pairwise system. The inferred interaction matrix is
\begin{equation}
    \hat A \approx
    \left( \begin{array}{rrr}
        0.06 & 3.91 & -0.09 \\
        -1.00 & -1.00 & -1.00 \\
        0.64 & -1.55 & 0.10
    \end{array} \right),
\end{equation}
and the inferred vector of intrinsic growth rates is $\hat{\mathbf{r}} \approx (-1.06,1.00,0.12)$.
Notably, species 3 now behaves as a predator of species 1, illustrating how the pairwise approximation can misrepresent the ecological relationships between species.
\medskip

\begin{figure}[t!]
\centering

\begin{tcolorbox}[
    enhanced,
    colframe=Header_color!80,   
    colback=Body_color!40,      
    coltext=black,
    width=\linewidth,
    arc=2mm,
    boxrule=0.4pt,
    left=6mm,
    right=6mm,
    top=3mm,
    bottom=3mm,
    fonttitle=\bfseries,
    coltitle=black,
    title=Box 1: Hessian matrices to detect HOIs
]
\justifying

\noindent
Starting from the general dynamics given by Eq.~\eqref{eq:Kolmogorov}, we compute the partial derivative of $\dot N_i$ with respect to $N_j$:
\begin{equation}
    J_{ij} (N_1, \ldots, N_n) = \frac{\partial \dot{N}_i}{\partial N_j} =
    N_i \frac{\partial F_i}{\partial N_j} + δ_{ij} F_i,
    \label{eq:jac}
\end{equation}
where $δ_{ij}$ is the Kronecker delta. Eq. (\ref{eq:jac}) measures how the growth rate of species $i$ responds to changes in the abundance of species $j$. Collecting all terms, we obtain the Jacobian matrix $J = (J_{ij})$ of the system \cite{strogatz2015}. Under the higher-order Lotka–Volterra dynamics given by Eq. (\ref{eq:hoLV}) for $n=3$, if $i \not= j$, then $J_{ij} = N_i (a_{ij} + b_{ipq} N_k)$, where $p = \mathrm{min}(j,k)$, $q = \mathrm{max}(j,k)$, and $k \not= i, j$, so that $J_{ij}$ depends also on species $k$ in the presence of higher-order interactions.

Computing the second partial derivative of $\dot N_i$ with respect to $N_j$ and $N_k$, we obtain
\begin{equation}
\begin{split}
    (H_{\dot N_i})_{jk} (N_1, \ldots, N_n) = \frac{\partial^2 \dot N_i}{\partial N_j \partial N_k} \\
    = N_i \frac{\partial^2 F_i}{\partial N_j \partial N_k} +
    δ_{ij} \frac{\partial F_i}{\partial N_k} +
    δ_{ik} \frac{\partial F_i}{\partial N_j},
\end{split}
\end{equation}
which measures how the growth rate of species $i$ responds to changes in the abundances of species $j$ and $k$. Collecting all terms, we obtain the Hessian matrix $H^{(i)} = H_{\dot N_i}$ associated with the function $\dot N_i$. Since we are assuming that $F_i$ is sufficiently smooth, this matrix is symmetric. Note that there is just one Jacobian matrix for the whole system, whereas there is one Hessian matrix for each species. Under the higher-order Lotka–Volterra dynamics given by Eq. (\ref{eq:hoLV}), $H^{(i)}_{jk} = H^{(i)}_{kj} = b_{ijk} N_i$ if $i$, $j$, and $k$ are distinct and $j<k$.

Therefore, in second-order Lotka-Volterra systems, higher-order interactions are absent if and only if all Hessian functions $H^{(i)}_{jk}$ vanish whenever $i$, $j$, and $k$ are distinct. By contrast, the Jacobian depends on both pairwise and higher-order contributions, and thus cannot be used to reliably detect higher-order interactions. Song and Saavedra \cite{song2021} present an example for $n = 3$ in which the Jacobian entries $J_{12} = N_1 (a_{12} + b_{123} N_3)$ and $J_{13} = N_1 (a_{13} + b_{123} N_2)$ change sign, and argue that such sign changes indicate the presence of higher-order interactions. While this is correct, the converse does not hold: higher-order interactions may be present without inducing sign changes in the Jacobian (see Fig.~\ref{fig:Counterex} for a counterexample).
\vspace{-0.25cm}
\begin{center}
\includegraphics[width=1\linewidth]{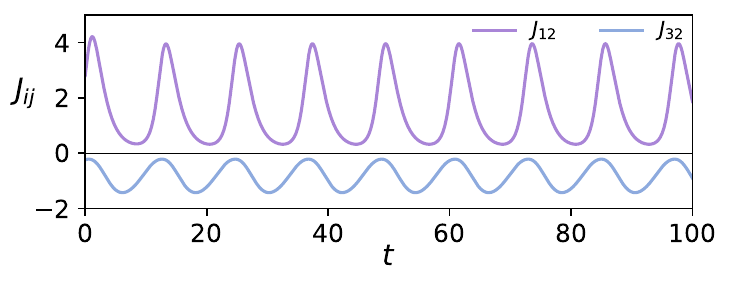}
\vspace{-0.85cm}\captionof{figure}{\small\justifying The Jacobian functions for System 2 are $J_{12}=N_1(4-1.2N_3)$, $J_{13}=-1.2N_1N_2$, $J_{21}=J_{23}=-N_2$, $J_{31}=1.2N_2N_3$, and $J_{32}=N_3(-3+1.2N_1)$. The figure shows that neither $J_{12}$ nor $J_{32}$ changes sign.}
\label{fig:Counterex}
\end{center}
\end{tcolorbox}
\end{figure}
To illustrate that this phenomenon is not restricted to minimal systems, we finally consider a more realistic community with seven species arranged across three trophic levels (System 3). Species 1 preys on species 2 and 3, species 2 preys on species 4 and 5, and species 3 preys on species 6 and 7. In addition, there is competition between species 2 and 3, between species 4 and 5, and between species 6 and 7.
All self-regulations coefficients are set to $-1$, and all intrinsic growth rates are set to 1.
The interaction matrix is
\begin{equation}
    A =
    \left( \begin{array}{lllllll}
    -1.0 & \phantom{-}5.3 & \phantom{-}4.1 & \phantom{-}0.0 & \phantom{-}0.0 & \phantom{-}0.0 & \phantom{-}0.0 \\
    -0.5 & -1.0 & -3.0 & \phantom{-}5.5 & \phantom{-}4.5 & \phantom{-}0.0 & \phantom{-}0.0 \\
    -1.1 & -3.8 & -1.0 & \phantom{-}0.0 & \phantom{-}0.0 & \phantom{-}5.4 & \phantom{-}5.8 \\
    \phantom{-}0.0 & -1.6 & \phantom{-}0.0 & -1.0 & -0.4 & \phantom{-}0.0 & \phantom{-}0.0 \\
    \phantom{-}0.0 & -1.5 & \phantom{-}0.0 & -0.6 & -1.0 & \phantom{-}0.0 & \phantom{-}0.0 \\
    \phantom{-}0.0 & \phantom{-}0.0 & -1.5 & \phantom{-}0.0 & \phantom{-}0.0 & -1.0 & -0.6 \\
    \phantom{-}0.0 & \phantom{-}0.0 & -1.6 & \phantom{-}0.0 & \phantom{-}0.0 & -0.6 & -1.0 \\
    \end{array} \right).
\end{equation}

We add three higher-order coefficients: $b_{123} = -4.9$, $b_{245} = -4.8$, and $b_{367} = -2.2$. The coefficient $b_{123}$ accounts for the negative effect that the competition between species 2 and 3 has on species 1, and similarly for $b_{245}$ and $b_{367}$ (see Fig. \ref{fig:Fig3}a). The relative weight of higher-order interactions is again maintained at $\theta \approx 0.18$. As in the previous system, the pairwise approximation is very accurate. The time series of abundances for all seven species are almost indistinguishable from the original ones (Fig. \ref{fig:Fig3}), with all correlation coefficients satisfying $\rho_i>0.99999$. Both dynamics stabilize to an almost identical limit cycle. On the other hand, the inferred interaction matrix is
{\setlength{\arraycolsep}{2pt}
\begin{equation}
\begin{array}{l}
\hat A \approx \\[2mm]
\left(
\begin{array}{rrrrrrr}
-1.13 & 9.31 & 4.27 & -229.18 & 245.51 & 76.21 & -66.45 \\
-0.54 & -0.68 & -2.97 & -7.81 & 15.78 & 3.22 & -3.27 \\
-1.03 & -4.49 & -1.22 & 35.68 & -39.23 & -7.26 & 14.38 \\
0.00 & -1.60 & 0.00 & -1.00 & -0.40 & 0.00 & 0.00 \\
0.00 & -1.50 & 0.00 & -0.60 & -1.00 & 0.00 & 0.00 \\
0.00 & 0.00 & -1.50 & 0.00 & 0.00 & -1.00 & -0.60 \\
0.00 & 0.00 & -1.60 & 0.00 & 0.00 & -0.60 & -1.00
\end{array}
\right),
\end{array}
\end{equation}}

\noindent and the inferred vector of intrinsic growth rates is $\hat{\mathbf{r}} \approx (-14.72,0.92,4.19,1.00,1.00,1.00,1.00)$. Beyond the changes in magnitude in the first three rows of $\hat A$, and the change in magnitude and sign in $\hat{r}_1$, the ecological interpretation of several interactions is altered. In particular, species 2 and 4 now appear as competitors, and the same happens for species 3 and 6 (see Fig. \ref{fig:Fig3}c for a schematic representation of the preserved and added interactions).

Our results reveal two qualitatively distinct scenarios. In System 1, higher-order interactions leave dynamical fingerprints that cannot be mimicked by any pairwise model. Their presence is, in principle, detectable. In Systems 2 and 3, by contrast, a fitted pairwise model closely tracks the higher-order dynamics both in the short and long term, but with misidentified interaction types and intrinsic growth rates. Moreover, as shown in the Supporting Information, there are systems in which the pairwise approximation closely matches higher-order trajectories over short time windows, yet diverges in their long-term behavior. Taken together, these scenarios define a taxonomy of failure modes for the inference of higher-order interactions from time series alone: higher-order interactions may be hidden in plain sight.
\bigskip

\noindent \textbf{\large Discussion}

\noindent Our results show that higher-order interactions (HOIs) can be effectively hidden in abundance dynamics. We identify systems in which, even when HOIs contribute substantially to the underlying mechanisms, their effects can be accurately reproduced by an approximated pairwise model. As a consequence, the presence or absence of HOIs cannot, in general, be inferred from time-series data alone. Our results present a fundamental limit of structural indistinguishability since different underlying mechanisms can generate nearly indistinguishable dynamics. Critically, the inferred pairwise model can assign qualitatively different ecological roles to species, including changes in sign of intrinsic growth rates, and changes in sign and direction of interactions. These results highlight the need to disentangle prediction from explanation. In the real presence of HOIs, a pairwise description may have excellent predictive power, but it is not explanatory. Occam’s razor does not simply instruct us to prefer pairwise models by default; rather, it reminds us that when HOIs are not identifiable from dynamics alone, simplicity buys predictive convenience but not necessarily explanatory truth.

Our approach is deliberately conceptual rather than methodological. We use a simple geometric lens on per capita growth rates (by viewing pairwise dynamics as trajectories constrained to hyperplanes, and HOIs as curvature in that space) to expose the limits of what abundance time series can reveal about interaction structure. This perspective clarifies why diagnostics based on the Jacobian, such as sign changes \cite{song2021}, are sufficient but not necessary to detect HOIs, as they can easily miss higher-order effects along the observed trajectories. Analytically, we show in Box 1 that for higher-order Lotka–Volterra systems the true mathematical fingerprint of HOIs resides in the Hessian of the growth-rate functions, not in the Jacobian alone.

Establishing the ecological reality of HOIs therefore requires going beyond time series. Our results advocate for coupling models with targeted empirical data, motivating the extension of experimental approaches used to estimate species interactions, such as response–surface designs \cite{hart2018quantify,inouye2001response}, beyond pairwise settings. However, such empirical approaches face a fundamental known limitation: the number of required experimental manipulations grows prohibitively large when considering interactions among pairs, triplets, and larger groups of species, making a fully systematic exploration impractical. Consistent with this challenge, recent work on epistatic higher-order interactions has shown that as the number of interacting species increases, the effect of the interactions becomes indistinguishable from noise \cite{llabres2026reducibility}.

Far from closing the door on HOI research, these limitations provide important new questions and directions. From a theoretical perspective, further work is needed to understand when and how higher-order interactions can be rigorously reduced to effective pairwise descriptions, as suggested in recent studies of complex systems \cite{llabres2026reducibility, peixoto2026graphs, lacasa2026equivalence}, and to clarify the ecological interpretation of such reductions. Indeed, while higher-order models intuitively offer greater descriptive power, their steep mathematical and computational costs complicate the application of established ecological frameworks, such as modern coexistence theory \cite{singh2021higher, majer2024higher} or local and structural stability \cite{chen2024stability, cenci2018structural}. Empirically, they suggest that to confidently claim HOIs are present, future research must incorporate life-history information, phenological data, biologically-justified non-additive terms, and other independent constraints on interaction structure that cannot be absorbed into effective pairwise terms \cite{wootton_putting_1994, case_testing_1981, buche_continuum_2025, letten_mechanistic_2019, mayfield_higher-order_2017, latremoliere2022, vandermeer1998,sundarraman_higher-order_2020, godoy2014phenology}. Ultimately, synthesizing these independent sources of information will answer whether we truly need a higher-order description for a given system.
\bigskip

\noindent \textbf{\large Acknowledgments}

\noindent V.C.S. acknowledges financial support from the Ministerio de Ciencia e Innovaci\'on (grant PID2021-127607OB-I00). S.L.O. acknowledges financial support from the Gobierno de Arag\'on through a doctoral fellowship. S.L.O., C.G.A. and J.G.G. acknowledge financial support from the Departamento de Industria e Innovaci\'on del Gobierno de Arag\'on y Fondo Social Europeo (FENOL group grant E36-23R) and from the Ministerio de Ciencia e Innovaci\'on (grant PID2023-147734NB-I00). S.M. acknowledges financial  support by the Spanish State Research Agency
(MICIU/AEI/10.13039/501100011033) and FEDER (UE) under projects COSASTI (PID2024-157493NB-C22), and the Mar{\'\i}a de Maeztu project CEX2021-001164-M.

\bibliography{references}

\end{document}